\documentclass[twocolumn]{article}
\textwidth=\paperwidth
\advance \textwidth by -1.5in
\textheight=\paperheight
\advance \textheight by -1.3in

\newlength{\abslength}
\abslength=\textwidth
\advance \abslength by -1in

\usepackage{graphicx}
\usepackage{amsmath}
\usepackage{amssymb}
\usepackage{enumerate}
\DeclareFontFamily{U}{rsfs}{\skewchar\font"7F}
\DeclareFontShape{U}{rsfs}{m}{n}{
	<-6> rsfs5
	<6-8> rsfs7
	<8-> rsfs10
	}{}
\DeclareMathAlphabet{\mathscr}{U}{rsfs}{m}{n}

\def\ii{{\mathrm{i}}}

\def\ee{{\mathrm{e}}}
\def\dd{{\mathrm{d}}}
\def\Re{\mathop{\mathrm{Re}}}
\def\Im{\mathop{\mathrm{Im}}}

\def\bracket#1{\langle #1 \rangle}

\def\sub#1{_{\mathrm{#1}}}

\def\Vec#1{\mbox{\boldmath $#1$}}

\def\4He{\mbox{$^4$He}}
\def\3He{\mbox{$^3$He}}

\def\etal{{\it et al. }}
\begin{document}

\baselineskip 12pt

\twocolumn[\begin{center}\noindent\begin{LARGE}{\bf Thermal dissipation in quantum turbulence}\end{LARGE}\\ \vspace{\baselineskip}Michikazu Kobayashi and Makoto Tsubota\\ \vspace{\baselineskip}Faculty of Science, Osaka City University, 3-3-138, Sumiyoshi-ku, Osaka 558-8585, JAPAN\\ \vspace{\baselineskip}\begin{minipage}{\abslength}

The microscopic mechanism of thermal dissipation in quantum turbulence is numerically studied by solving the coupled system involving the Gross-Pitaevskii equation and the Bogoliubov-de Gennes equation.
At low temperatures, the obtained dissipation does not work at scales greater than the vortex core size.
However, as the temperature increases, dissipation works at large scales and it affects the vortex dynamics.
We successfully obtain the mutual friction coefficients of the vortex in dilute Bose-Einstein condensates dynamics as functions of temperature.

\vspace{\baselineskip}
%{\bf KEYWORDS: }
\end{minipage}\vspace{\baselineskip}\end{center}]

The physics of quantum turbulence (QT), comprising tangled quantized vortices, is one of the most important research topics in low-temperature physics \cite{Donnelly}.
Stimulated by recent experiments on both superfluid \4He and superfluid \3He where a few similarities were observed between quantum and classical turbulences \cite{Maurer,Stalp,Bradley}, studies on QT have entered a new stage where one of the main motivations is to investigate the relationship between quantum and classical turbulences.

With this motivation, we theoretically investigated QT in our previous studies by numerically solving the Gross-Pitaevskii (GP) equation \cite{Kobayashi-1, Kobayashi-2}.
Since the GP equation is applicable to compressible quantum fluids, compressible excitations of wavelengths smaller than the vortex core size, which affect the vortex dynamics in QT, are emitted during vortex reconnections, the disappearance of small vortices, or by high frequency Kelvin waves.
As a result, the intrinsic behavior of quantized vortices in QT, such as the Richardson cascade, is hindered by the compressible short-wavelength excitations.
To eliminate these excitations, we introduced a phenomenological dissipation term that works only at scales smaller than the vortex core size and successfully obtained the Kolmogorov law in QT, which is one of the most important statistical laws in classical turbulence \cite{Frisch}.
However, the microscopic origin of dissipation and the realistic nature of the introduced dissipation term in quantum fluids were still unknown.

In the research field of superfluid \4He, the dissipation mechanism in quantized vortices has been studied for approximately 50 years with the consequent development of the concept of mutual friction.
The hydrodynamics of superfluid \4He is usually described using the two-fluid model; in this model, the system consists of an inviscid superfluid and viscous normal fluid, and dissipation in quantized vortices is caused by the mutual friction between the vortices and normal fluid \cite{Hall}.
Although mutual friction is very important in QT \cite{Vinen-1,Tough,Schwarz,Barenghi,Finne}, no microscopic theory exists that can reproduce the mutual friction coefficients as functions of temperature.
This is because the superfluid \4He system is strongly correlated and extremely difficult to study microscopically by using, for example, the quantum field theory.

Quantized vortices were also discovered in systems of atomic Bose-Einstein condensates (BECs) \cite{Madison,Abo-Shaeer}.
Since the systems are different from superfluid \4He, they can be easily studied by using the GP equation for the condensate and the Bogoliubov-de Gennes (BdG) equation for the excitations from the condensate; these equations should also prove useful in revealing the microscopic dissipation mechanism in quantized vortices.
However, in the research field of atomic BECs, no concept exists that explains dissipation or mutual friction of quantized vortices.
Some groups \cite{Choi,Kasamatsu,Zaremba} have discussed dissipation in atomic BECs; however, they have not directly considered the dissipation mechanism in quantized vortices.
In this work, we are the first to study the dissipation mechanism in quantized vortices in dilute BECs.
With this study, we cannot only justify the phenomenological dissipation term introduced in our previous work but also define the mutual friction in a dilute BEC by considering its relationship with the mutual friction in superfluid \4He.

It would be fairly reasonable to assume that dissipation in quantum fluids, including the quantized vortices is caused by the interaction between the condensate and its excitations.
The dynamics of the excitations can be described by the BdG equation; therefore, we numerically solve the time development of the coupled system involving the GP and BdG equations for a quantum fluid with some quantized vortices.
Our results reveal that dissipation obtained by the coupled equations can work only at small scales and reproduces the phenomenological dissipation term introduced at very low temperatures, as mentioned in our previous studies.
As the temperature increases, however, dissipation works at larger scales and it affects the vortex dynamics; this is qualitatively similar to the mutual friction of quantized vortices in superfluid \4He.
Moreover, our model is limited to a dilute Bose gas.
Thus, we further investigate dissipation as a model for mutual friction in dilute BECs by calculating the dynamics of one straight vortex, and we successfully obtain the friction coefficients as functions of temperature.
It is impossible to directly compare the obtained friction coefficients with those of superfluid \4He because both the GP and BdG equations are applicable only to dilute BECs.
However, this study can be regarded as the first study elucidating the mutual friction of quantized vortices in dilute BECs.

To consider a quantum fluid as a Bose-Einstein condensed system, we start with the many-body Hamiltonian,
\begin{equation} \hat{H} = \int \dd \Vec{x} \: \hat{\Psi}^\dagger \Bigg[-\nabla^2 - \mu + \frac{g}{2}|\hat{\Psi}|^2 \Bigg] \hat{\Psi}. \label{eq-hamiltonian} \end{equation}
Its dynamics can be described by
\begin{equation} \ii \frac{\partial \hat{\Psi}}{\partial t} = [ - \nabla^2 - \mu + g \hat{\Psi}^\dagger \hat{\Psi} ] \hat{\Psi}. \label{eq-boson-field-dynamics} \end{equation}
Here, $\hat{\Psi}(\Vec{x},t)$ is the boson field operator, $\mu$ the chemical potential, and $g$ the coupling constant.
In the Bose-Einstein condensed system, the field operator $\hat{\Psi}(\Vec{x},t)$ can be represented in terms of the mean-field ansatz \cite{Girardeau,Castin},
\begin{equation} \hat{\Psi} = \Phi + \hat{\chi} + \hat{\zeta}, \label{eq-mean-field} \end{equation}
which conserves the particle number.
Here, we define the terms appearing in Eq. (\ref{eq-mean-field}): the macroscopic wave-function is represented as $\Phi(\Vec{x},t) = \mathcal{O}(\sqrt{N_0 / V})$, the first-ordered excitations as $\hat{\chi}(\Vec{x},t) = \mathcal{O}(1 / \sqrt{V})$, the higher-ordered excitations as $\hat{\zeta}(\Vec{x},t) = \mathcal{O}(1 / \sqrt{N_0 V})$, the number of condensate particles as $N_0$, and the volume of the system as $V$.
Substituting Eq. (\ref{eq-mean-field}) into Eq. (\ref{eq-boson-field-dynamics}) and neglecting the higher-ordered excitations $\hat{\zeta}(\Vec{x},t)$, we obtain the GP equation as
\begin{equation} \ii \frac{\partial \Phi}{\partial t} = [ - \nabla^2 - \mu + g (|\Phi|^2 + 2 \bracket{\hat{\chi}^\dagger \hat{\chi}})] \Phi + g \bracket{\hat{\chi}^2} \Phi^\ast \label{eq-GP} \end{equation}
and the BdG equation \cite{Bogoliubov,Gennes} as
\begin{equation} \ii \frac{\partial \hat{\chi}}{\partial t} = [- \nabla^2 - \mu + 2g|\Phi|^2] \hat{\chi} + g \Phi^2 \chi^\dagger. \label{eq-BdG} \end{equation}
When the macroscopic wave-function is expressed as $\Phi(\Vec{x},t) = f(\Vec{x},t) \ee^{\ii \phi(\Vec{x},t)}$, $f(\Vec{x},t)^2$ is considered to be the condensate density, and the superfluid velocity $\Vec{v}(\Vec{x},t)$ is given by $\Vec{v}(\Vec{x},t) = 2 \nabla \phi(\Vec{x},t)$.
The vorticity $\Vec{\omega}(\Vec{x},t) = \nabla \times \Vec{v}(\Vec{x},t)$ vanishes everywhere in a single connected region of the fluid and thus any rotational flow is carried only by quantized vortices.
In the core of each vortex, $\Phi(\Vec{x},t)$ vanishes; therefore, the circulation $\oint\Vec{v} \cdot \Vec{\dd s}$ around the core is quantized by $4\pi$.
The vortex core size is given by the healing length $\xi = 1 / f \sqrt{g}$.

The GP equation (\ref{eq-GP}) can be expressed as $\ii \partial \Phi(\Vec{x},t) / \partial t = H\sub{GP} \Phi(\Vec{x},t)$.
The Hamiltonian $H\sub{GP}$ of the GP equation has the following imaginary term:
\begin{equation} \Im[H\sub{GP}] = - \gamma = g\Im\Bigg[\frac{\bracket{\hat{\chi}^2} \Phi^\ast}{\Phi} \Bigg]. \label{eq-gamma} \end{equation}
This defines the dissipation $\gamma(\Vec{x},t)$ of the condensate caused by the interaction with the noncondensed particles.
We can quantitatively calculate the dissipation $\gamma(\Vec{x},t)$ from Eq. (\ref{eq-gamma}) by numerically solving the coupled system involving the GP equation (\ref{eq-GP}) and BdG equation (\ref{eq-BdG}).

\begin{figure}[htb]
\centering \includegraphics[width=0.8\linewidth]{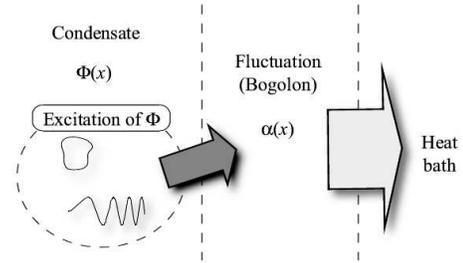}
\caption{\label{fig-energy-flow} Image of the energy flow in a system of quantum fluid coupled with a heat bath.} \end{figure}

To solve the BdG equation (\ref{eq-BdG}), we use the Bogoliubov transformation; therefore,
\begin{equation} \hat{\chi} = \frac{1}{\sqrt{V}} \sum_{j} [u_j \hat{\alpha}_j + v_j^\ast \hat{\alpha}_j^\dagger], \label{eq-Bogoliubov-transformation} \end{equation}
where $u_j(\Vec{x},t)$ and $v_j(\Vec{x},t)$ are the Bogoliubov coefficients, and $\hat{\alpha}_j$ and $\hat{\alpha}^\dagger_j$ are the annihilation and creation operators of a quasiparticle, respectively.
We assume that the quasiparticles are coupled with a heat bath at temperature $T$; further, we use the local equilibrium assumption
\begin{equation} \bracket{\hat{\alpha}^\dagger_j \hat{\alpha}_j} = N_j = \frac{1}{\exp(E_j / T)-1} \label{eq-local-equilibrium} \end{equation}
with the excitation spectrum $E_j$ of quasiparticles.
The local equilibrium assumption (\ref{eq-local-equilibrium}) yields the energy flow in the system, as shown in Fig. \ref{fig-energy-flow}.
When excitations such as vortices or sound waves are formed in the condensate, their energy is transferred to quasiparticles and finally dissipated to the heat bath.
By using Eqs. (\ref{eq-Bogoliubov-transformation}) and (\ref{eq-local-equilibrium}), we can deduce the final form of the coupled system involving the GP and BdG equations:
\begin{subequations} \label{eq-final-GP}
\begin{align} \ii \frac{\partial \Phi}{\partial t} = {} & [- \nabla^2 - \mu +g (|\Phi|^2 + 2 n\sub{e})] \Phi + g m\sub{e} \Phi^\ast, \\ \ii \frac{\partial u_j}{\partial t} = {} & [- \nabla^2 - \mu + 2 g |\Phi|^2] u_j - g \Phi^2 v_j = A_j, \\ \ii \frac{\partial v_j}{\partial t} = {} & - [- \nabla^2 - \mu + 2 g |\Phi|^2] v_j + g \Phi^{\ast 2} u_j = B_j, \\ n\sub{e} = {} & \sum_j[|u_j|^2 N_j + |v_j|^2 (N_j + 1)], \\ m\sub{e} = {} & - \sum_j[u_j v_j^\ast (2 N_j + 1)], \\ E_j = {} & \int \dd \Vec{x} \: \Re[u_j^\ast A_j + v_j^\ast B_j]. \end{align}
\end{subequations}

\begin{figure}[htb] \centering \begin{minipage}{0.49\linewidth} \centering \includegraphics[width=0.99\linewidth]{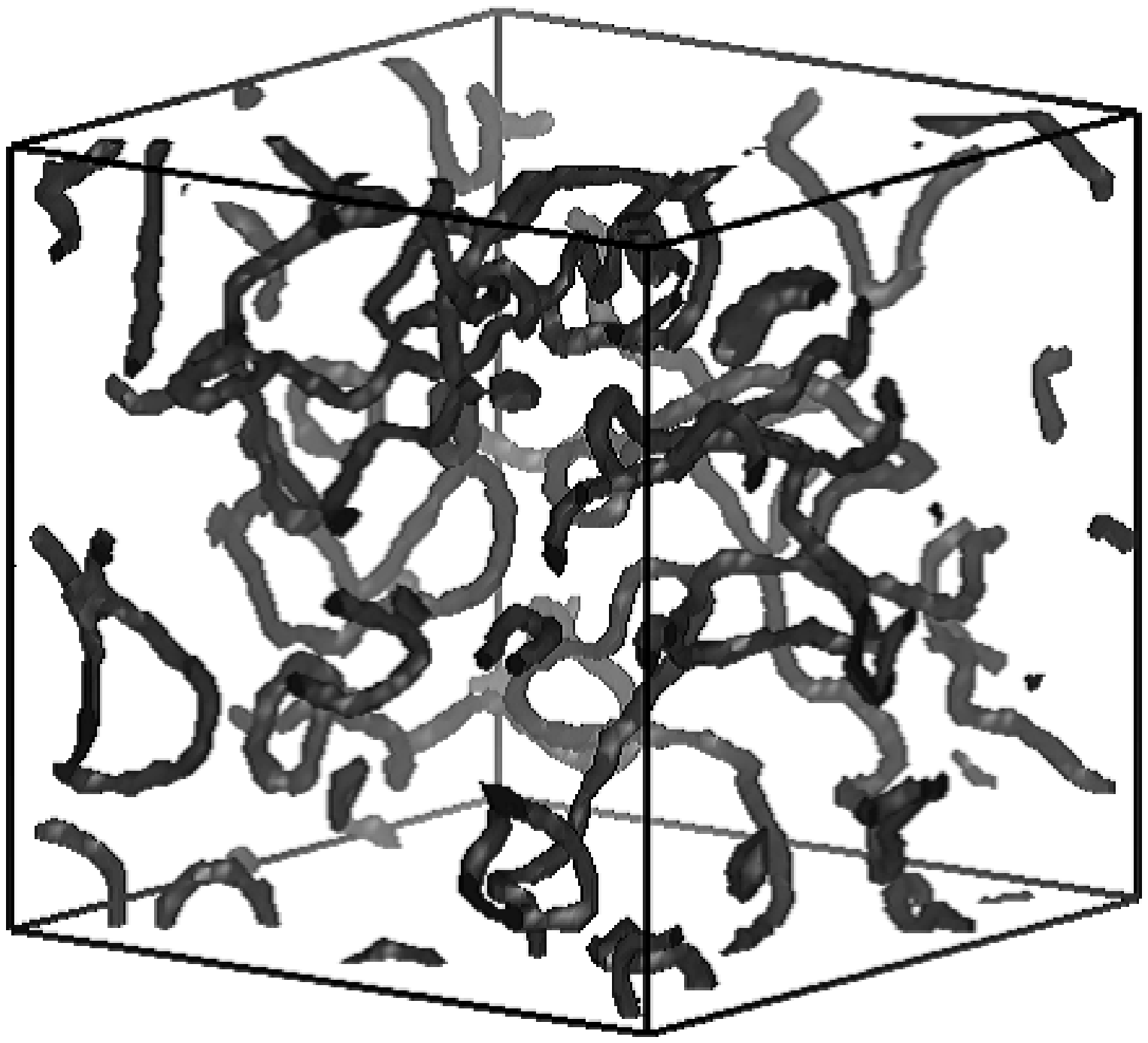} \\ (a) \end{minipage} \begin{minipage}{0.49\linewidth} \centering \includegraphics[width=0.99\linewidth]{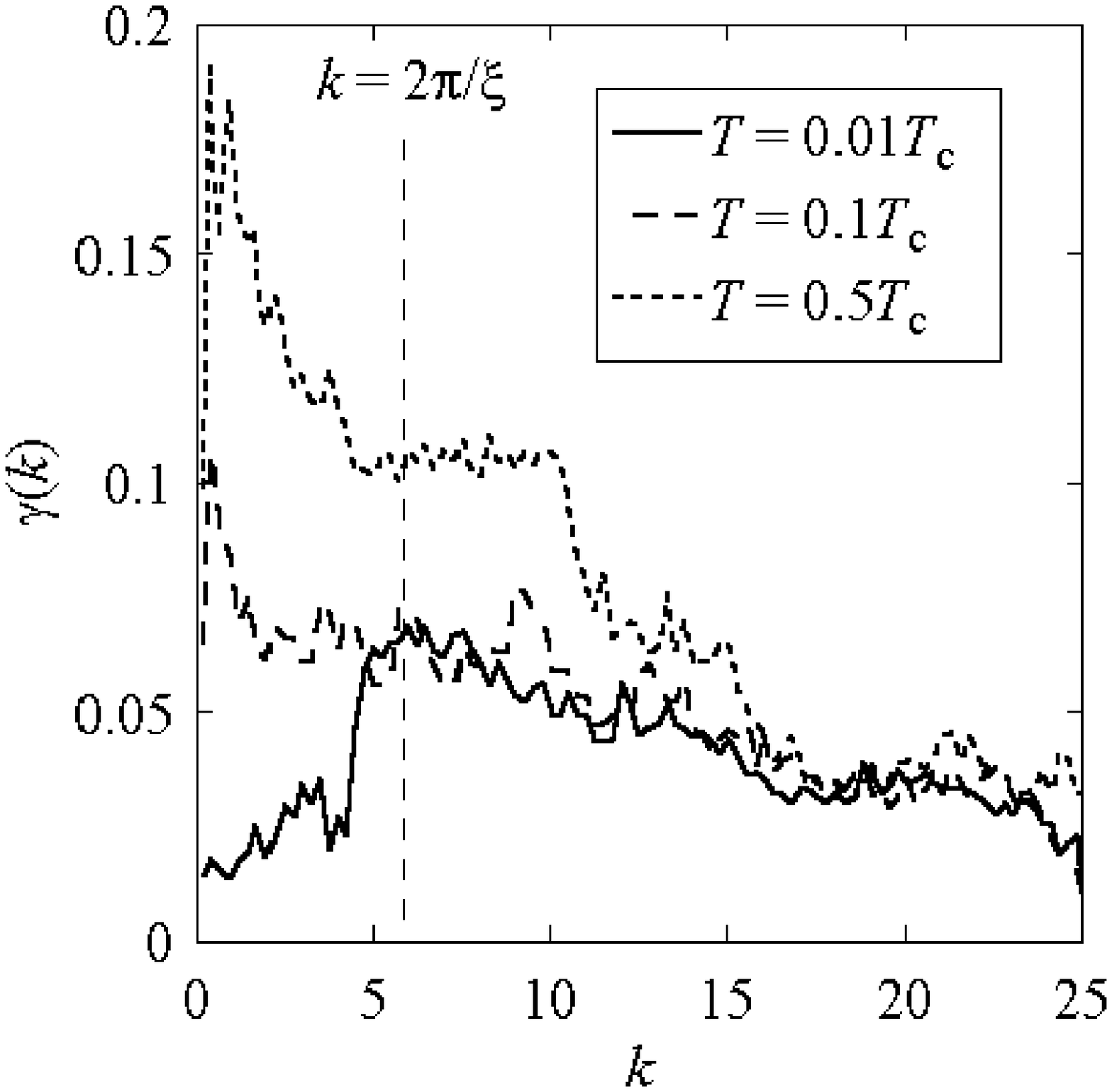} \\ (b) \end{minipage}
\caption{\label{fig-dissipation} (a) An example of the configurations of quantized vortices at $t = 0$. (b) Wave number dependence of the Fourier-transformed dissipation term $\gamma(k,t = 1)$, which is obtained by performing an ensemble average of 25 initial states.} \end{figure}

We numerically solve the coupled equations (\ref{eq-final-GP}).
We use a pseudospectral method \cite{Press} in space with periodic boundary conditions in a box with a spatial resolution of $32^3$ grid points.
With regard to the numerical parameters, assuming $g = 1$, we use a spatial resolution of $\Delta x = 0.125$ and $V = 4^3$, where the length scale is normalized by $\xi$.
The numerical time evolution is given by the Runge-Kutta-Gill method \cite{Press} with a temporal resolution of $\Delta t = 1 \times 10^{-4}$.
We begin with a macroscopic wave-function $\Phi(\Vec{x},t=0)$ that includes several randomly placed vortices, as shown in Fig. \ref{fig-dissipation} (a), and with the uniform excitations $\hat{\chi}(\Vec{x},t=0)$ given by,
\begin{align} u_j(\Vec{x},t=0) = {} & \ee^{\ii \Vec{k}_j \cdot \Vec{x}} \sqrt{\frac{1}{2V} \frac{k_j^2 + g |\Phi|^2}{E_j} + 1}, \nonumber \\ v_j(\Vec{x},t=0) = {} & \ee^{- \ii \Vec{k}_j \cdot \Vec{x}} \sqrt{\frac{1}{2V} \frac{k_j^2 + g |\Phi|^2}{E_j} - 1}, \label{eq-initial-fluctuation} \end{align}
where $k_j = 2 \pi j/ \sqrt[3]{V}$, where $j$ is an integer.
Subsequently, we calculate the dissipation term $\gamma(\Vec{x},t)$ at $t = 1$.
Figure \ref{fig-dissipation} (b) shows the dependence of the Fourier-transformed dissipation $\gamma(k,t)$ on the wavenumber $k$ at several temperatures: $T = 0.01 T\sub{c}$, $0.1 T\sub{c}$, and $0.5 T\sub{c}$, where $T\sub{c} = 4 \pi / \{\zeta(3/2)\}^{2/3}$ is the critical temperature for the Bose condensation of free bosons.
At a low temperature of $T = 0.01 T\sub{c}$, dissipation works only at wave numbers greater than $2 \pi / \xi$, which is consistent with the dissipation term $\gamma(k) = \gamma_0 \theta(k - 2 \pi / \xi)$ with a step function $\theta$ introduced in our previous studies \cite{Kobayashi-1,Kobayashi-2}.
From this result, we find that only short-wavelength excitations emitted during the vortex reconnections, the disappearance of small vortices, or by high frequency Kelvin waves get dissipated at scales smaller than the vortex core size.
This phenomenon has also been confirmed by another simulation that starts from a state without vortices and obtains smaller and less $k$-dependent values of $\gamma(k,t)$.
On the other hand, as the temperature increases, dissipation works at small wave numbers as well.
Since dissipation at small waven umbers dissipates vortices at scales greater than the vortex core size $\xi$, the vortex dynamics are directly affected by this dissipation.
Hence, we can expect an effect of $\gamma(\Vec{x},t)$ similar to that of mutual friction in superfluid \4He.

\begin{figure}[htb] \centering \begin{minipage}{0.49\linewidth} \centering \includegraphics[width=0.99\linewidth]{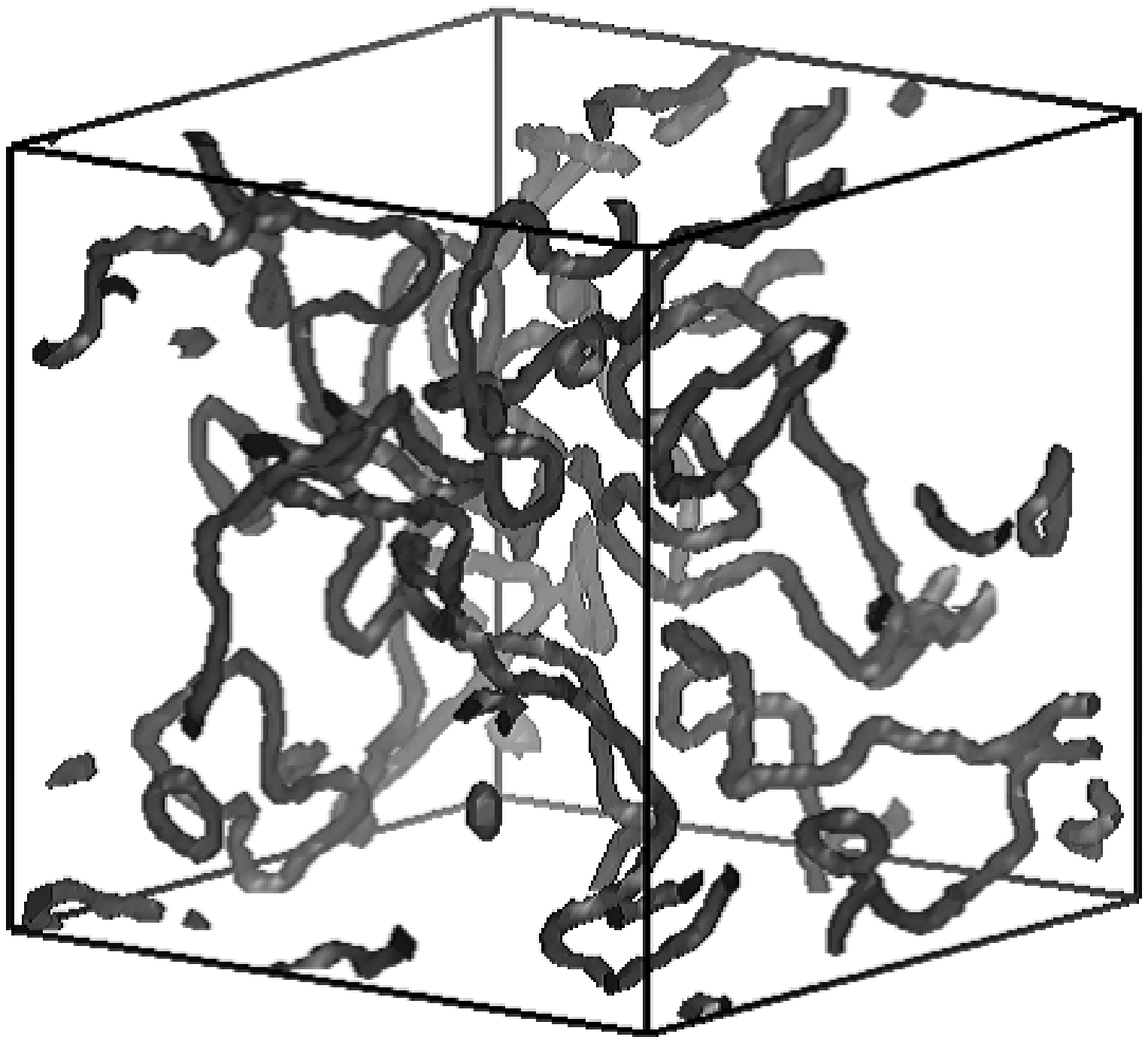} \\ (a) \end{minipage} \begin{minipage}{0.49\linewidth} \centering \includegraphics[width=0.99\linewidth]{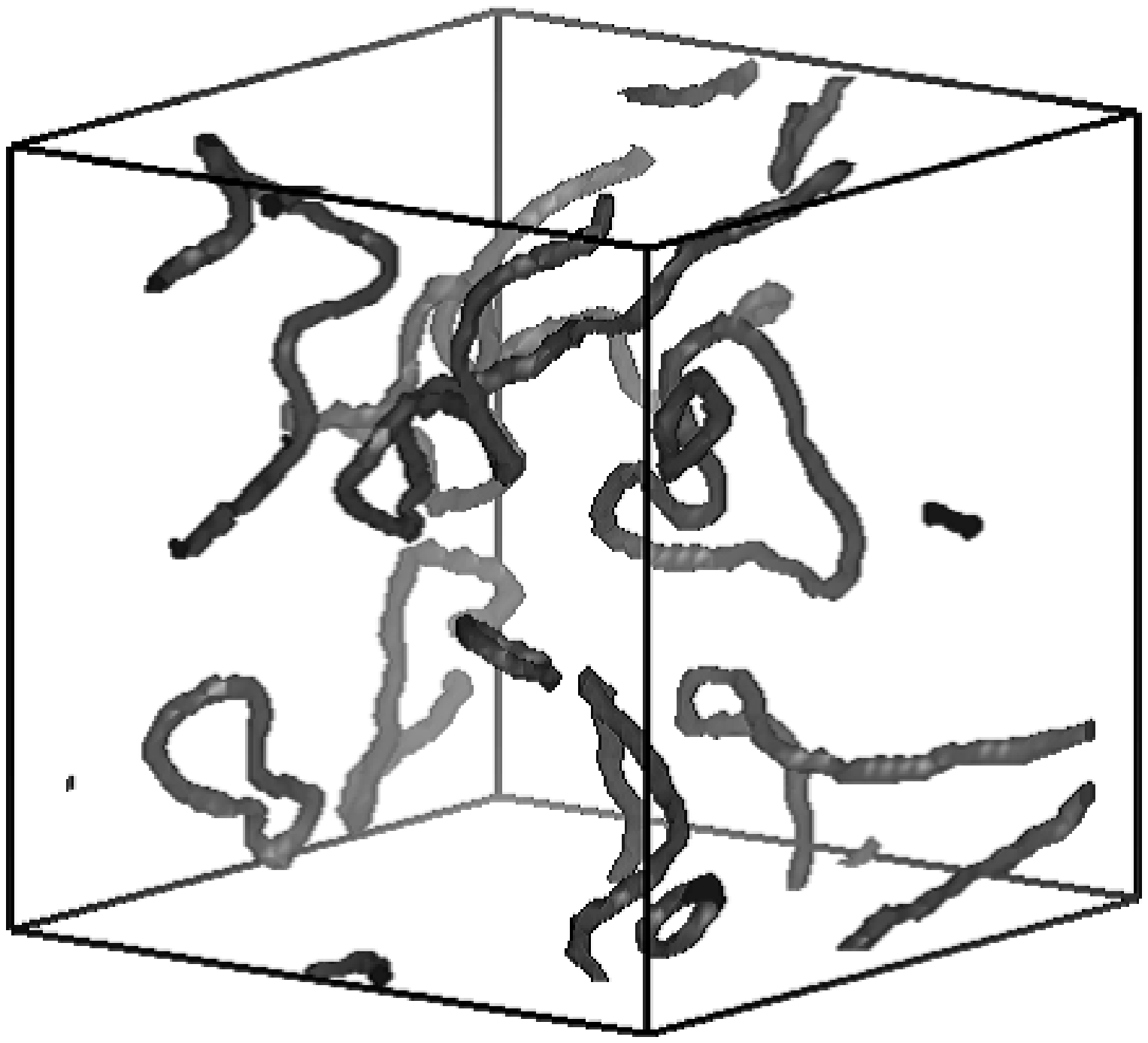} \\ (b) \end{minipage}
\caption{\label{fig-vortex-configuration} Configurations of quantized vortices at $t = 1$ at $T = 0.01 T\sub{c}$ (a) and $T = 0.1 T\sub{c}$ (b), starting from the configuration shown in Fig. \ref{fig-dissipation} (a).} \end{figure}

Figures \ref{fig-vortex-configuration} (a) and \ref{fig-vortex-configuration} (b) show the vortex configurations at $t = 1$ and for $T = 0.01 T\sub{c}$ and $T = 0.1 T\sub{c}$ respectively, starting from the vortex configuration shown in Fig. \ref{fig-dissipation} (a).
Because dissipation at large scales affects both small vortices and short-wavelength Kelvin waves along the vortices, the vortex configurations clearly show the difference in the effects on vortex dynamics at different temperatures.
We can observe fewer vortices and fewer Kelvin waves in Fig. \ref{fig-vortex-configuration} (b) than in Fig. \ref{fig-vortex-configuration} (a), which obviously conforms to the effect of dissipation at large scales.
By using the vortex-filament model, Tsubota \etal investigated the difference in the vortex configurations in QT at different temperatures and observed fewer Kelvin waves at higher temperatures \cite{Araki}.
Their observation is qualitatively consistent with our result.

\begin{figure}[htb] \centering  \includegraphics[width=0.5\linewidth]{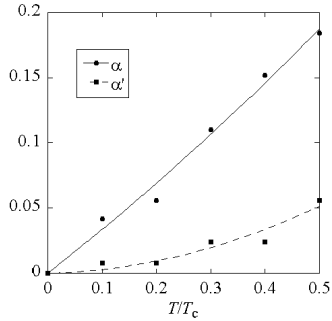}
\caption{\label{fig-mutual-friction} Temperature dependence of the coefficients $\alpha$ and $\alpha^\prime$. Plots represent the numerical results and lines indicate some fitting.} \end{figure}

Next, we attempt to calculate the coefficients of mutual friction as functions of temperature.
When one straight vortex along the $z$-axis is placed under the velocity field $\Vec{v}\sub{e} = [v\sub{e}, 0, 0]$, the dynamics of the vortex position $\Vec{s}(t) = [s_x(t), s_y(t), 0]$ can be described by mutual friction as $\dot{\Vec{s}}(t) = [\alpha^\prime v\sub{e}, \alpha v\sub{e}, 0]$ \cite{Schwarz,Hall}.
Here, $\alpha$ and $\alpha^\prime$ are the coefficients that characterize the amplitude of mutual friction.
The solution becomes
\begin{equation} \Vec{s}(t) = [s_x(0) + \alpha^\prime v\sub{e} t, s_y(0) + \alpha v\sub{e} t, 0]. \label{eq-mutual-friction-solution} \end{equation}
Starting from the state with one straight quantized vortex, we numerically solve the coupled equation involving the GP and BdG equations under the velocity field
\begin{subequations}
\begin{align} \ii \frac{\partial \Phi}{\partial t} = {} & [- \nabla^2 - \mu + g (|\Phi|^2 + 2 n\sub{e}) + \ii \Vec{v}\sub{e} \cdot \nabla] \Phi \nonumber \\ & + g m\sub{e} \Phi^\ast, \\ \ii \frac{\partial u_j}{\partial t} = {} & [- \nabla^2 - \mu + 2 g |\Phi|^2 + \ii \Vec{v}\sub{e} \cdot \nabla] u_j - g \Phi^2 v_j, \\ \ii \frac{\partial v_j}{\partial t} = {} & - [- \nabla^2 - \mu + 2 g |\Phi|^2 + \ii \Vec{v}\sub{e} \cdot \nabla] v_j \nonumber \\ & + g \Phi^{\ast 2} u_j, \end{align}
%\begin{align} \ii \frac{\partial \Phi}{\partial t} = {} & [- \nabla^2 - \mu + g (|\Phi|^2 + 2 n\sub{e}) + \ii \Vec{v}\sub{e} \cdot \nabla] \Phi + g m\sub{e} \Phi^\ast, \\ \ii \frac{\partial u_j}{\partial t} = {} & [- \nabla^2 - \mu + 2 g |\Phi|^2 + \ii \Vec{v}\sub{e} \cdot \nabla] u_j - g \Phi^2 v_j, \\ \ii \frac{\partial v_j}{\partial t} = {} & - [- \nabla^2 - \mu + 2 g |\Phi|^2 + \ii \Vec{v}\sub{e} \cdot \nabla] v_j + g \Phi^{\ast 2} u_j, \end{align}
\end{subequations}
for the case of $v\sub{e} = 0.1$.
We can calculate $\alpha$ and $\alpha^\prime$ by comparing the position of the vortex in the numerical simulation using Eq. (\ref{eq-mutual-friction-solution}).
Figure \ref{fig-mutual-friction} shows the temperature dependence of $\alpha$ and $\alpha^\prime$.
The effect of the dissipation term $\gamma(\Vec{x},t)$ on the vortex dynamics at large scales and its monotonic increase with temperature are evident, which is qualitatively consistent with mutual friction in superfluid \4He at temperatures much lower than the superfluid critical temperature.
This result develops the first estimation of mutual friction of quantized vortices in dilute BECs.
The temperature dependence of $\alpha$ and $\alpha^\prime$ needs to be experimentally observed, and it may become a standard scale for measuring the temperature in atomic BECs with quantized vortices.

This work allows quasiparticles to move via Eq. (\ref{eq-final-GP}), but we do not discuss the motion here.
With regard to turbulence in superfluid \4He, Vinen predicted that at finite temperatures, superfluid and normal fluid are likely to be coupled together at large scales due to mutual friction and thus behave similarly to the turbulence in a one-component fluid \cite{Vinen-2}.
In our simulation, we can also expect a similar coupled turbulence in which the dynamics of the quasiparticles is strongly coupled with that of the condensate with both the dynamics becoming comparable at large scales.
This physics will be reported shortly.

In conclusion, we investigated the dissipation mechanism in quantized vortices by numerically solving the coupled equation involving the GP and BdG equations.
At low temperatures, dissipation works at scales smaller than the vortex core size, which is consistent with the phenomenological dissipation term introduced in our previous studies.
As the temperature increases, dissipation can work at large scales as well, and it directly affects the vortex dynamics.
We successfully related this effect to the mutual friction in superfluid \4He by calculating the mutual friction coefficients as functions of temperature.

We thank W. F. Vinen for the useful discussions.
MT acknowledges the supports of Grant in Aid for Scientific Research from JSPS (Grant No. 18340109) and Grant in Aid for Scientific Research on Priority Areas (Grant No. 17071008) from MEXT.

\end{document}